\documentstyle[11pt,newpasp,twoside]{article}
\def\edcomment#1{\iffalse\marginpar{\raggedright\sl#1\/}\else\relax\fi}
\reversemarginpar

\begin{document}
\title{The Discovery of Submillimeter Galaxies}
\author{A.\,W. Blain}
\affil{Department of Astronomy, Caltech 105-24, Pasadena CA91125, USA} 

\begin{abstract}
I briefly describe some results about 
luminous distant dusty galaxies obtained in the 5 years since 
sensitive two-dimensional bolometer array cameras became available. 
The key requirements for making additional 
progress in understanding the properties 
of these galaxies are discussed, especially the potential 
role of photometric redshifts based on radio, submillimeter (submm) 
and far-infrared(IR) continuum observations
\end{abstract}

\section{Introduction}

In 1997 the SCUBA camera was commissioned at the 
15-m JCMT in Hawaii, providing 2.5-arcmin-wide images at wavelengths 
of 450 and 850\,$\mu$m with resolutions of 
7 and 15\arcsec respectively. The MAMBO 1.25-mm camera at the 30-m 
IRAM telescope provides a similar capability (see Dannerbauer et al. 2002), 
while the 350-$\mu$m SHARC-II and 1.1-mm BOLOCAM cameras being 
commissioned at the 10-m Caltech Submillimeter Observatory (CSO) should  
provide significantly enhanced performance within a year. 
In the future, BOLOCAM is 
expected to observe from the 50-m mm-wave GTM/LMT telescope under construction 
in Mexico, a larger-format 
8$\times$8-arcmin$^2$ camera SCUBA-II is being designed in the UK, and  
developments of MAMBO are planned for APEX -- a new 12-m telescope at 
the 5000-m ALMA site in Chile. APEX will join the 10-m-class Japanese ASTE 
submm telescope at the ALMA site for which a large bolometer camera is 
being designed. 

SCUBA was the first submm camera able to survey fields large 
enough to detect  
the redshifted thermal dust emission from previously unknown galaxies 
(Smail, Ivison 
\& Blain 1997). The 
peak of the emission from galaxies, typically at 60--200\,$\mu$m in 
their restframe, 
corresponding to a range of 
dust temperatures between about 60 and 20\,K,
is redshifted into SCUBA's observing bands. The steep 
submm Rayleigh--Jeans slope of 
the dust emission ensures that distant galaxies with similar bolometric
luminosities and spectral energy distributions (SEDs) 
would produce similar flux densities at all redshifts from 
$z \simeq 0.5$ to $z \sim 5$,
assuming the same spectral 
energy distribution (SED). The detectability of
galaxies does depend on the details of the SED, in general being greater 
for cooler temperatures at a fixed luminosity and redshift  
(see Blain et al.\ 2002). 

More than 300 high-$z$ submm galaxies (SMGs) have now 
been detected by SCUBA and 
MAMBO, while BOLOCAM (Glenn et al.\ 1998) could detect new examples at 
a rate of order 1 per hour. Various types of surveys have been made:
in narrow fields to exploit the gravitational lensing effect of 
clusters of galaxies (Smail et al.\ 2002; Chapman et al.\ 2002a; 
Cowie, Barger \& 
Kneib 2002); in the Hubble Deep Field 
(Hughes et al.\
1998); and in wider shallower surveys (Eales et al.\ 1999; Borys et al.\ 
2002; Scott et al.\
2002) covering a total of about 0.25\,deg$^2$.  

These SMGs are responsible for a significant fraction of the 
star-formation/ AGN activity at $z>1$ (Blain et al.\ 1999a). 
It is vital to understand their individual properties if we are to understand 
the formation of galaxies as a whole. Their inferred space density is 
similar to that of giant Elliptical galaxies in the local Universe, 
and it has been suggested that they are high-$z$ formation events of 
these rare galaxies (Eales et al.\ 1999), 
presumably in the most massive dark-matter halos
with the lowest specific angular momenta. It is more likely that they 
reflect a more common, short-lived phase involving the formation of 
galactic bulges in
perhaps episodic mergers (Blain et al.\ 1999b). The test of these ideas 
is to measure the redshifts, clustering properties and mass distribution 
of a representative sample of SMGs. 

\section{Finding redshifts and studying astrophysics} 
The coarse ($\sim$15\arcsec) resolution of submm images ensures that 
there are many possible faint optical counterparts to the detected 
galaxy. Hence, while some SMGs have bright optical counterparts  
(Ivison et al.\, 1998, 2000), most  
remain unidentified at optical wavelengths. Final confirmation of 
a correct identification in both position and redshift 
is provided by detecting (sub)mm-wave CO molecular line 
emission with a suspected redshift in the narrow spectral 
window of existing (sub)mm-wave spectrographs
(Frayer et al.\ 1998, 1999). 

Mm-wave continuum interferometer images of the 
fields have reduced positional uncertainties to $\sim 1$\,arcsec for 
several SMGs (Downes et al.\ 1999; Frayer et al.\ 2000; Dannerbauer 
et al.\ 2002), but are very expensive in observing time. 
Deep near-IR imaging to $K > 22$ tends to reveal plausible red 
counterparts for many SMGs (Smail et al.\ 1999; Frayer et al.\ 
2003) by their $J-K$ and $I-K$ colors. Deep optical/near-IR spectroscopy 
of these galaxies can then be attempted to find redshifts. The most efficient 
technique for determining redshifts, however, appears to be to exploit 
very deep radio images. For reasonable SEDs, SMGs with 850-$\mu$m 
flux densities of about 5--10\,mJy 
(with luminosities $\sim 10^{13}$\,L$_\odot$) should be detectable 
in $\sim 10$\,$\mu$Jy RMS 1.4-GHz VLA images out to $z \simeq 3$,  
if they  
lie on the far-IR--radio correlation observed for local galaxies 
(Condon 1992): 
(see Barger, Cowie \& Richards (2000) and Chapman et al.\, (2001, 2002b). The 
wide field (0.25\,deg$^2$) and accurate sub-arcsec astrometry of  
VLA images, coupled with the low surface density of the faintest 
radio sources as compared with optically-selected galaxies yield accurate 
positions for a large fraction ($\sim 70$\%) of 
the SMGs brighter than 5\,mJy at 850\,$\mu$m. This provides an 
opportunity for efficient multi-object optical spectroscopy. 
In March 2002, about 25 spectra were obtained using the Keck-LRIS 
spectrograph 
(Chapman et al.\ 2003). These
will be subject to CO molecular line spectroscopy to confirm 
the identifications 
and to study both their gas masses (via velocity dispersions) and excitation 
conditions  
(via line--line and line--continuum ratios), using  
mm-wave interferometers in the 2002-2003 Northern winter. 
Only a handful of 
reliable redshifts were available for SMGs in 2001: now it is 
likely that a luminosity function of these  
radio-selected SMGs should be available 
in 2003. 

\section{Photometric redshifts} 

The key target of investigating the SMGs is now to find their 
physical properties, especially their masses. However, just 
obtaining a reasonably complete redshift distribution is important for 
fixing their form of evolution and ensuring their fractional contribution 
to the energy emission of all galaxies is correctly accounted for 
(Eales et al.\ 1999; Blain et al.\ 1999a). 

As redshift surveys were generally unsuccessful  
until (Chapman et al.\ 2003), photometric techniques have been proposed to 
provide redshift information (Eales et al.\ 1999; Hughes et al.\ 
1998, 2002). The key information available is 
radio (Carilli \& Yun 1999), 
submm and mid-/far-infrared photometry, typically from VLA, 
SCUBA and {\it ISO} respectively. The {\it IRAS} survey is not 
sufficiently deep to detect SMGs; {\it ISO} 
data is deeper, but covers only a small area and is only useful for 
the closest (Soucail et al.\
1999) or brightest examples (Ivison et al.\ 1998). 
The {\it SIRTF} space telescope will be
ideal for finding far-IR SEDs.

Fitting a thermal spectrum to a galaxy at uncertain redshift leads to 
an unavoidable degeneracy between the inferred dust temperature $T$ 
and redshift $z$. The peak wavelength of the SED is determined in the 
observers frame, but 
this is shifted in exactly the same way by either 
a fractional increase in $T$
or a corresponding fractional decrease in $(1+z)$. This makes any 
far-IR/submm-based photometric redshift only as accurate as  
the knowledge of the temperature (Blain, Barnard \& Chapman 2003), and 
not to $\Delta z \simeq 0.5$ as claimed by Hughes et al.\ (2002). 
Despite non-thermal radio emission being due to an entirely different 
process, the submm--radio properties of galaxies on the far-IR--radio 
correlation conspire to produce a similar 
$T$--$(1+z)$ 
degeneracy if $T<60$\,K (Blain 1999). 
If a reliable link exists between dust temperature and luminosity, then 
it is possible to break this degeneracy; however, the accuracy of the 
result is then determined by the scatter in the LT relation, which is 
likely to be at least 30\%, implying at least this great an uncertainty 
in redshift (Blain et al.\ 2003). 

The addition of $K$-band near-IR 
data (Dannerbauer et al.\ 2002) could also help, but first the intrinsic 
scatter in the ratio between the $K$-band and far-IR luminosity of the galaxies 
must be known. Based on observations of fairly 
complete samples of SMGs (Ivison et al.\ 1998; Frayer et al.\ 2003), 
the K-band magnitudes are certainly scattered by as much as 
$\Delta K \sim 2$\,mag. 
Spectroscopic redshifts remain essential for accurate study of SMGs. 

\section{Spectroscopic mm-wave redshifts} 

Correct identifications of SMGs via either samples of faint radio galaxies 
that feed targets to multi-object spectrographs or 
deep near-IR imaging and 
spectroscopy, must be confirmed and verified using mm-wave 
interferometers or single-antenna 
telescopes to detect CO molecular rotational 
line emission at integer multiples of 
115\,GHz in the galaxy's rest frame. 

There are other possibilities for obtaining spectroscopic redshifts, 
especially the direct detection of molecular or atomic fine-structure  
spectral lines 
at mid-/far-IR and submm wavelengths. The key is to obtain 
wide-band spectral coverage at these wavelengths, in order to allow 
searches for redshifts. At present, mm-wave spectrographs 
cover a total frequency range of only 
$\Delta \nu/\nu \simeq 0.05$, and so a redshift must 
be known to well within 5\% before confirmation can be made. 

Powerful correlators for the 100-m GBT will allow searches for CO(1-0) lines 
(rest frequency 115\,GHz) from high-redshift galaxies in the 22 \& 44-GHz
radio bands. At shorter submm wavelengths new very stable 230/345-GHz 
spectral line receivers at 
the CSO will have $\sim 10$\,GHz bandwidths (Rice et al. in prep.). 
Dispersive techniques may 
allow very-wideband, low-resolution spectrographs at millimeter or 
far-infrared wavelengths. These include the ZSPEC and WaFIRS 
waveguide/grating concepts for 
space-borne and ground-based applications (Bradford et al. in prep). 

\acknowledgments

I thank Vicki Barnard, Matt Bradford, Scott Chapman, Dave Frayer, 
Andrew Harris, Tom Phillips, Naveen Reddy, Frank Rice, Chip Sumner 
and Jonas Zmuidzinas. Work on a wide-band correlator for the CSO is 
being supported by the Caltech Tolman endowment and the Research 
Corporation.  

%\section{References} 


\begin{references} 
Barger, A.J., Cowie, L.L., \& Richards E.A., 2000, \aj, 119, 2092\\
Blain, A.W. 1999, \mnras, 309, 955\\
Blain, A.W., Barnard, V.E., \& Chapman, S.C. 2003, \mnras, 338, 733\\
Blain, A.W., Smail, I., Ivison, R.J., \& Kneib, J.-P., 1999a, \mnras, 
302, 632\\
Blain, A.W. et al. 
1999b, \mnras, 209, 715\\ 
Blain, A.W., Smail, I., Ivison, R.J., Kneib, J.-P., \& Frayer, D.T. 2002, 
Physics Reports, in press (astro-ph/0202228)\\
Borys, C., Chapman, S.C., Halpern, M., \& Scott, D. 2002, \mnras, 330, 63\\
Carilli, C.L., \& Yun, M.S. 1999, \apj, 513, L13\\
Chapman, S.C., et al. 2001, \apj, 548, L147\\
%Chapman, S.C., Richards, E.A., Lewis G.F., Wilson G., \& Barger, A.J. 2001, 
%\apj, 548, L147\\
Chapman, S.C., Scott, D., Borys, C., \& Fahlman, G.G. 2002a, \mnras, 330, 92\\
Chapman, S.C., et al. 2002b, \apj, in press\\
Chapman, S.C., et al. 2003, Nature, in press\\
Condon, J.J., et al. 1992, ARA\&A, 30, 575\\ 
Cowie, L.L., Barger, A.J., \& Kneib, J.-P., 2002, \aj, 123, 2197\\
Dannerbauer, H., et al. 
2002, \apj, in press (astro-ph/0201104)\\
Downes, D. et al. 1999, \aap, 347, 809\\
Eales, S., et al. 1999, \apj, 515, 518\\ 
Frayer, D.T., et al. 1998, \apj, 506, L7\\
Frayer, D.T., et al. 1999, \apj, 514, L13\\
Frayer, D.T., Smail, I., Ivison, R.J., \& Scoville, N.Z. 2000, \aj, 120, 1668\\
Frayer, D.T., et al. 2003, \aj, submitted\\
Glenn, J., et al. 1998, Proc. SPIE, 3357, 326\\ 
%Holland, W.S., et al. 1999, \mnras, 303, 659\\ 
Hughes, D.H., et al. 1998, Nature, 394, 241\\
Hughes, D.H., et al. 2002, \mnras, 335, 871\\ 
Ivison, R.J., et al. 1998, \mnras, 298, 583\\
Ivison, R.J., et al. 2000, \mnras, 315, 209\\
Scott, S.E., et al. 2002, \mnras, 331, 817\\
Smail, I., Ivison, R.\,J., \& Blain, A.\,W. 1997, \apj, 490, L5\\
Smail, I., et al. 1999, \mnras, 308, 1061\\ %ERO
Smail, I., et al. 2002, \mnras, 331, 495\\ 
Soucail, G., et al. 1999, \aap, 343, L70\\
%Soucail, G., Kneib, J.-P., B\'ezecourt, J., Metcalfe, L., Altieri, B. \& 
%Le Borgne, J.-F. 1999, \aap, 343, L70\\
\end{references}
\end{document}